%&Plain TeX
\magnification=1200

\hsize    = 160 true mm %length of line
\vsize    = 200 true mm  %vertical length of text
\parskip  = 3 true  pt plus 1 true pt minus 1 true  pt
\baselineskip = 24 true pt plus 1 true pt minus 1 true pt 
\def\cen{\centerline}

\def\endproof{\nobreak\kern5pt\nobreak\vrule height4pt width4pt depth0pt}
\def\eth{{\mathord{\mkern8mu\bf\mathaccent"7020{\mkern-8mu\partial}}}}
\def\barm{{\overline m}}
% % % % % % % % % % % % Insert text below: % % % % % % % % % % % % % % % % %
%\special{!userdict begin /bop-hook{gsave 10 2000 translate
%270 rotate /Times-Roman findfont 144 scalefont setfont
%0 0 moveto 0.8 setgray (DRAFT DRAFT DRAFT) show grestore}def end}

\noindent
\cen{\bf Bounded Area Theorems for Higher Genus Black Holes}
\smallskip
\cen{\it E. Woolgar}
\cen{Dept.~of Mathematical Sciences}
\cen{and}
\cen{Theoretical Physics Institute}
\cen{University of Alberta}
\cen{Edmonton, AB}
\cen{Canada T6G 2G1}
\cen{e-mail: ewoolgar@math.ualberta.ca}
\bigskip
\noindent
{\bf Abstract.} 
By a simple modification of Hawking's well-known topology theorems 
for black hole horizons, we find lower bounds for the areas of smooth
apparent horizons and smooth cross-sections of stationary black hole 
event horizons of genus $g>1$ in four dimensions. For a negatively 
curved Einstein space, the bound is ${{4\pi (g-1)}\over {-\ell}}$ 
where $\ell$ is the cosmological constant of the spacetime. This is 
complementary to the known upper bound on the area of $g=0$ black holes 
in de Sitter spacetime. It also emerges that $g>1$ quite generally 
requires a mean negative energy density on the horizon. The bound is 
sharp; we show that it is saturated by certain extreme, asymptotically 
locally anti-de Sitter spacetimes. Our results generalize a recent 
result of Gibbons.

\par\vfill\eject

\cen{\bf Bounded Area Theorems for Higher Genus Black Holes}
\bigskip
\cen{\bf I. Introduction}
\smallskip
\noindent
Black holes embedded in ``locally adS'' background spacetimes (backgrounds 
locally isometric to spacetimes of constant negative curvature; adS = 
anti-de Sitter) have been seminal to recent developments in black hole 
physics. Most notably, these black holes have been essential to recent 
progress in understanding black hole entropy, but they have forced us to 
revisit other issues as well. For example, in contrast to the situation 
in asymptotically flat spacetimes, static anti-de Sitter black holes can 
have horizons of non-zero genus. This observation has led us to improve 
our understanding of topological censorship [1] and its implications for 
black hole horizon topology [2].

Related to the locally adS black holes are the ``asymptotically locally 
adS'' solutions of Mann [3], and Brill, Louko, and Peld\'an [4]. These 
are vacuum, negative scalar curvature spacetimes with topology $R^2\times
\Sigma_g$, where $\Sigma_g$ is a Riemann surface of genus $g$. This integer 
$g$ and a real number $m$ referred to as the {\it mass} (or {\it energy}) 
parametrize the family of solutions. The term ``mass'' reflects the fact 
that $m$ is a conserved charge conjugate to a Killing vector field that 
is timelike at infinity. For $m$ less than a {\it negative} 
minimum value $m_0$, there are no horizons and the solutions are nakedly 
singular. For $m>m_0$, future event horizons exist and for each such value
of $m$ and every $g\ge 2$, one can construct a not-nakedly-singular, 
maximally extended spacetime, {\it even for $m<0$}. The zero-mass solution 
is everywhere locally isometric to adS spacetime, but is also a black hole 
and has trapped surfaces. The $m=m_0$ case has Killing horizons that are
not event horizons, so this is a nakedly singular spacetime, not a black
hole. We will refer to this case as an ``extreme solution'' rather than
as an extreme black hole.

The identification of $m$ as mass is contentious. Vanzo [5] gives explicit 
mass formul\ae\ both for the special class of solutions under discussion 
here and for more general stationary black holes with the same asymptotic 
behaviour. At issue is the zero of mass-energy, since $m$ can be negative. 
Vanzo advocates shifting the zero of energy so that the total energy is
$m-m_0$, assuming there is suitable shift $m_0$ that works not only for
the special solutions considered here and by Vanzo, but for all reasonable
solutions with the same asymptotic behaviour. Contrary-minded, the parameter
$m$ appears be the {\it gravitational} mass; the tidal deformations of rings
of particles are those produced by a source whose gravitational mass is $m$.
This is evident from the form of the Weyl spinor, which is type II--II:
$$\Psi_{ABCD}=kr^{-3}\alpha_{(A}\alpha_B\beta_C\beta_{D)}\eqno{(1)}$$
for $r$ an ``area radial coordinate'' and $k$ a numerical factor 
times the mass, whence the geometry is conformally flat iff $m=0$.

Still this interpretation of $m$ is startling, for assuming the 
{\it gravitational} weak equivalence principle ({\it cf}. [6]), then we
lose the positive mass theorem. This is not completely unexpected, however. 
The generator of the time translation symmetry here 
is not the usual ``timelike rotation generator'' ($J_{04}$ in a common
notation) of the anti-de Sitter group; it's a ``boost-like generator'' 
($J_{34}$), and this undermines a key step in the Witten-type derivations 
such as that of [7].\footnote{$^1$}
{It is also true that the topological identifications needed to 
compactify the horizon create obstructions to the global Killing spinors
that enter parts of the Witten-type arguments and that generate
supersymmetries. Solutions of the Witten equation may equally encounter 
global obstructions. However, even if this were not the case, the energy
defined by $J_{34}$ would not yield to Witten-type positivity arguments.}

Nonetheless, in the class of not-nakedly-singular solutions discussed
above, and the slightly generalized class of electrovac solutions discussed
below, the mass is bounded below, and very recently a general boundedness 
argument has been given [8] for solutions that share this asymptotic 
behaviour. This is reminiscent of a conjecture of Horowitz and Myers [9] 
made in a different (but perhaps not unrelated) context and suggests that 
a physical mechanism acts to stabilize the theory and prevent arbitrarily 
negative energies.

Herein we obtain lower bounds on the area of certain embedded surfaces,
including apparent horizons and cross-sections of stationary black hole 
event horizons, with genus $g>1$. While these computations are perhaps 
most relevant to black holes in adS backgrounds, the results apply more 
generally. The next section contains the derivations. Our main result
follows from a variation within a spacelike hypersurface and is applicable 
to both apparent horizons and stationary event horizons. We also briefly
discuss a variation within a null hypersurface of an event horizon 
cross-section. The discussion section shows that the area bound is 
saturated by higher genus event horizons of certain extreme solutions. 
In that section, we also suggest that the area bound plays a role in the 
stability of solutions with higher genus horizons.

As well as being applicable to both apparent horizons and stationary 
event horizons, our main result does not require the hypersurface in 
which the variation takes place to be a moment of time symmetry, a
maximal slice, or otherwise special. We note, however, that apparent 
horizons lying in hypersurfaces that are moments of time symmetry are 
minimal surfaces. The area bound for them is then a standard result of 
hyperbolic (Riemannian) geometry [10,11] (but beware [10] erroneously 
omits a factor of 2 in the final result). Soon after the results of 
this paper were obtained, the author became aware of a preprint by 
Gibbons [8] independently deriving the area bound in this special case. 
The more general proof given below may well have been known to Gibbons 
when [8] was written since it is a variation on an argument given in 
[12], wherein the idea is attributed to Gibbons.
Finally, the results herein are closely related to an argument of [13]
asserting an {\it upper} bound for black hole horizon area in positively
curved backgrounds.

\bigskip
\cen{\bf II. Bounded Area Theorems}
\smallskip
\noindent
Let $\Sigma$ denote a compact, orientable, 
spacelike 2-surface embedded in a spacelike hypersurface $H$, which 
itself is embedded in spacetime $({\cal M},g_{ab})$. We require $\Sigma$
to be smooth, marginally outer trapped (the outgoing null geodesics 
orthognal to it must have zero divergence), and without outer trapped 
surfaces lying outside it in $H$. Apparent horizons and, under certain 
global assumptions, cross-sections of stationary event horizons have 
the last two of these properties, but should be noted that event 
horizons need not be smooth [14] and that it is not known whether 
apparent horizons are always smooth. The first fundamental form (or 
Riemannian metric induced by $g_{ab}$) on $H$ is $h_{ab}$, while the 
first fundamental form on $\Sigma$ is $\gamma_{ab}$.

The derivation is a modification of an argument given in [12]. As this 
is not an easily-obtained standard reference, we provide details. We will 
rely heavily on the text [15], and therefore will use the sign conventions 
of that text.

We fix a complex null frame at every point of $\Sigma$ as follows. 
First, let $l^a$ and $n^a$ be real and orthogonal to $\Sigma$, with 
$l^a$ future-outward directed, $n^a$ future-inward directed, and 
$l^an_a=1$. This still leaves the overall freedom $l^a\mapsto e^wl^a$, 
$n^a\mapsto e^{-w}n^a$, with $w:\Sigma\to R$. Fix $w$ by requiring that 
$${1\over {\sqrt{2}}}\left ( l^a-n^a\right ) =:r^a\in TH\quad ,
\eqno{(2)}$$
so that $r^a$ is the unit outward-directed normal to $\Sigma$ in $H$.
The remaining two elements of the tetrad, $m^a$ and $\barm^a$, are 
tangent to $\Sigma$ and obey $m_a\barm^a=-1$. 

For some {\it a priori} arbitrary function $y:\Sigma\to R$, we define
$$v^a=e^yr^a\in T\Sigma\quad ,\eqno{(3)}$$
and use it to construct a variation $\Sigma(r)$, $r\in[0,1]$, of 
the surface $\Sigma=\Sigma_0$ by pushing each point $p$ of $\Sigma$ a 
parameter distance $\nu$ along the geodesic of $h_{ab}$ whose initial 
point is $p$ and whose initial tangent vector is $v^a|_p$. We will later 
choose $y$ so that $\Sigma_{\nu}$ is a particularly useful variation of 
$\Sigma$, but for now it will remain unspecified. We may extend 
$m^a$ and $\barm^a$ to complex null fields tangent to $\Sigma_{\nu}$.
To extend $l^a$ and $n^a$ to null fields orthogonal to $\Sigma_{\nu}$,
we need to impose the hypersurface orthogonality conditions
$$\eqalignno{
l_a\left ( m^b v^a{}_{;b}-v^bm^a{}_{;b}\right ) &=0\quad ,&({\rm 4a)}\cr
n_a\left ( \barm^b v^a{}_{;b}-v^b\barm^a{}_{;b}\right ) &=0\quad .
&({\rm 4b)}\cr
}$$

Following [12], we will use the Newman-Penrose spin coefficient formalism.
The formalism is explained in Chapter 4 of [15].
By giving the quantities appearing in (4) their names in this formalism,
the hypersurface orthogonality conditions may be written as
$$\eqalignno{
\kappa-\tau-\delta y +{\overline \alpha}+\beta&=0\quad ,&({\rm 5a})\cr
\nu-\pi+{\overline \delta}y+\alpha+{\overline \beta}&=0\quad .
&({\rm 5b})\cr
}$$
These are equations (7.1) and (7.2) of [12], respectively.

The Newman-Penrose quantity $\rho:=m^a{\overline m}^b\nabla_bl_a$ is real,
in virtue of Prop.~(4.14.2) of [15], and represents the convergence of 
the geodesic congruence tangent at $H$ to the $l^a$ field. We seek to 
determine the change in $\rho$ along the variation defined above:
$${{d\rho}\over {dr}} := v^a(\rho)={{e^y}\over {\sqrt{2}}}\left ( D\rho
-D'\rho\right )\quad .\eqno{(6)}$$
The derivatives $D\rho:=l^a\nabla_a\rho$ and $D'\rho:=n^a\nabla_a\rho$ 
are given by the Newman-Penrose equations (4.11.12a) and (4.11.12f)
of [15]:
$$\eqalignno{
D\rho=&\rho\left ( \rho+\epsilon+{\overline \epsilon}\right ) 
+\sigma{\overline \sigma}+\Phi_{00}-{\overline \kappa}\tau -\kappa
\left ( 3\alpha+{\overline \beta}-\pi\right ) +{\overline \delta}\kappa
\quad ,& ({\rm 7a})\cr
D'\rho=&-\rho{\overline \mu}-\sigma\lambda-\tau{\overline \tau}+\kappa\nu
+\rho\left ( \gamma+{\overline \gamma}\right )-\tau 
\left ( \alpha -{\overline \beta}\right ) - \Psi_2-2\Lambda
+{\overline \delta}\tau\quad .&({\rm 7b})}$$
Here we have used the ``normalized spin frame conditions'' $\beta'=-\alpha$,
$\alpha'=-\beta$, and $\Pi=\Lambda$, together with the fact that $\delta'
={\overline \delta}$. As well, so that our notation follows that of 
[12], we use the symbols $\pi$, $\lambda$, $\mu$, and $\nu$ for 
$-\tau'$, $-\sigma'$, $-\rho'$, and $-\kappa'$ respectively. 

If we plug expressions (7) into (6) and simplify using (5), we obtain
$$\eqalign{
{{d\rho}\over {dr}} =& {{e^y}\over {\sqrt{2}}}\bigg \{
\rho \left ( \rho+\epsilon+{\overline \epsilon}
-\gamma-{\overline \gamma}+{\overline \mu}\right )+\sigma{\overline
\sigma}+\lambda\sigma+\left ( \kappa-\tau\right )\left ( {\overline {
\kappa-\tau}}\right )\cr
&+\Phi_{00}+\Psi_2+2\Lambda+\eth\delta y -{\overline \eth}\left ( 
{\overline \alpha}+\beta\right ) \bigg \} \quad ,\cr}\eqno{(8)}$$
where $\eth:=\delta-{\overline \alpha}+\beta$. This is equation (7.3)
of [12] (except for an inconsequential sign difference in the 
$\rho{\overline \mu}$ term).

Most terms appearing above have simple interpretations in terms of 
familiar geometrical quantities. For example, $\rho$, $\sigma$, and 
$\mu$ are related to the null extrinsic curvatures of $\Sigma$, while
$\Lambda$, $\Phi_{00}$, and $\Psi_2$ are built from components of the 
spacetime Riemann tensor. The Gauss curvature of $\Sigma$ is simply 
twice the real part ({\it cf}. Prop.~(4.14.21) of [15]) of the combination
$$K:=-\lambda\sigma-\Psi_2+\rho\mu+\Phi_{11}+\Lambda\quad ,\eqno{(9)}$$
where $K$ is known as the {\it complex curvature} of $\Sigma$.
Moreover, the term $\eth\delta y$ is simply the Laplacian $\Delta y$ in 
$(\Sigma,\gamma_{ab})$. Hence we can write
$$\eqalign{
{{d\rho}\over {dr}} =& {{e^y}\over {\sqrt{2}}}\bigg \{
\rho \left ( \rho+\epsilon+{\overline \epsilon}-\gamma-
{\overline \gamma}+\mu+{\overline \mu}\right )+\sigma{\overline \sigma}
+\left ( \kappa-\tau\right )\left ( {\overline { \kappa-\tau}}\right )\cr
&+\Delta y +\Phi_{00}+\Phi_{11}+3\Lambda-K-{\overline \eth} \left ( 
{\overline \alpha}+\beta\right ) \bigg \}\quad .\cr}\eqno{(10)}$$
Every term in this expression is manifestly real except for the last two,
whose imaginary parts must therefore cancel (this is a manifestation of 
Prop.~(4.14.43) of [15]).

Finally, we must specify the precise nature of the variation vector 
field $v^a$, or in other words we must specify $y$. We use the trick 
introduced by Hawking, which was to note that if $\int_{\Sigma}f(x)
d^2\Sigma=0$ then one can always solve $\Delta y +f(x)=0$ on $\Sigma$. 
That is, we choose $y$ to solve
$$\Delta y + \rho \left ( \rho+\epsilon+{\overline \epsilon}-\gamma-
{\overline \gamma}+\mu+{\overline \mu}\right )+\Phi_{00}+\Phi_{11}
+3\Lambda-K-{\overline \eth} \left ( {\overline \alpha}+\beta\right ) 
=c\quad ,\eqno{(11)}$$
for a constant $c$ to be determined below. Then (10) becomes simply
$${{d\rho}\over {dr}} = {{e^y}\over {\sqrt{2}}}\left [
\sigma{\overline \sigma} +\left ( \kappa-\tau\right )
\left ( {\overline { \kappa-\tau}}\right ) + c\right ]\quad .\eqno{(12)}$$
The constant $c$ is evaluated by integrating (11) over $\Sigma$. When doing
so, note that the $\eth$-term appearing there is a divergence, so it does
not contribute to the integral. Also, $\rho$ is zero on $\Sigma$ by
assumption. Moreover, the real part of $K$ is just one-half the Gauss 
curvature, so the integral of this term gives $\pi$ times the Euler 
characteristic of $\Sigma$, and for a Riemann surface of genus $g$ that 
characteristic is $2(1-g)$. Therefore,
$$cA(\Sigma)=\int_{\Sigma}\left ( \Phi_{00}+\Phi_{11}+3\Lambda\right )
d^2\Sigma + 2\pi(g-1)\quad ,\eqno{(13)}$$
where $A(\Sigma)$ is the area of $\Sigma$.

Hawking proved his topology theorem by next observing that $c\le 0$, for
if the first-order change in $\rho$, as given by (12), were positive
everywhere on $\Sigma$ and since $\rho |_{\Sigma}=0$, then the variation 
$\Sigma(r)$ would be an outer trapped surface just outside $\Sigma$, 
contrary to assumption. He then imposed an energy condition so that 
the integral on the right of (13) was non-negative, whence he concluded 
that $g=0,1$. We will reverse this last step and argue instead that the 
non-positivity of $c$ forces that integral term to be negative whenever 
$g>1$. That is, defining
$$\langle E\rangle={{2\int_{\Sigma}\left ( \Phi_{00}+\Phi_{11}+3\Lambda
\right ) d^2\Sigma}\over {A(\Sigma)}}\quad ,\eqno{(14)}$$
then (13), with $c\le 0$, gives
$$\eqalignno{
&\langle E \rangle A(\Sigma)+4\pi(g-1)\le 0\quad ,&({\rm 15a})\cr
\Rightarrow\quad &\langle E \rangle\le{{4\pi(1-g)}\over {A(\Sigma)}}<0
{\rm \quad for\ }g>1\quad ,&({\rm 15b})\cr
\Rightarrow\quad &A(\Sigma)\ge {{4\pi(g-1)}\over {-\langle E \rangle}}
\quad .&({\rm 15c})\cr}$$
The division in (15c) makes sense whenever $g>1$ in virtue of 
(15b).

From the Einstein equations, which in this signature are
$$R_{ab}-{1\over 2}g_{ab}R+\ell g_{ab}=-8\pi T_{ab}\quad ,\eqno{(16)}$$
for stress-energy tensor $T_{ab}$ and cosmological constant $\ell$,
we have ({\it cf}. equation (4.6.32) of
[15])
$$\eqalignno{
\Phi_{ab}=&4\pi \left ( T_{ab}-{1\over 4}g_{ab}T\right ) \quad ,
&({\rm 17a})\cr
\Lambda=&{{\pi}\over 3} T +{1\over 6}\ell\quad ,&({\rm 17b})\cr
\Rightarrow\quad E=&8\pi T_{ab}l^a \left ( l^b+n^b \right ) +\ell
\quad .&({\rm 17c})\cr}$$
For $T_{ab}=0$, then $\langle E \rangle=\ell$ so (15c) yields the bound
quoted in the Abstract.

An area bound can also be obtained by taking the hypersurface $H$
to be null and $\Sigma$ to be its intersection with a future event 
horizon, by a simple modification of the derivation quoted in [16] 
of Hawking's topology theorem for event horizons. Since [16] is a 
standard reference, we will quote from it, {\it mutatis mutandis}. 
We therefore use the sign conventions of [16] from here to the end 
of this section. These are generally opposite to those of [15] used 
in the rest of this article.

Using properties of stationary event horizons, it is shown in [16] that 
the right-hand side of their equation (9.7) cannot be positive. Following
their notation, let $\partial{\cal B}(\tau)$ denote an event horizon
cross-section ``at time $\tau$.'' The Einstein equation appearing below 
(9.7) of [16] must be corrected by adding the cosmological constant $\ell$ 
to the right-hand side. Let ${\tilde E}:=8\pi T_{ab}Y^a_1Y^b_2+\ell$, 
where $Y^a_{1,2}$ respectively denote in [16] the fields 
called $l^a$ and $n^a$ above (so they are null and normalized such 
that $Y^a_1Y_{2a}=-1$, using the signature employed in [16]). Let
$\langle {\tilde E}\rangle$
be the mean of ${\tilde E}$ over the surface ${\partial{\cal B}(\tau)}$:
$$\langle {\tilde E}\rangle 
=8\pi{{\int_{\partial {\cal B}(\tau)} T_{ab}Y^a_1Y^b_2dS}\over 
{A(\partial {\cal B}(\tau))}}+\ell\quad ,\eqno{(18)}$$ 
where $A(\partial {\cal B}(\tau))$ is the area of the cross-section
$\partial {\cal B}(\tau)$.

By substituting this expression into (9.7) of [16] and using the sign 
constraint subsequently deduced therein (the argument parallels that 
used to constrain the sign of $c$ above),\footnote{$^2$} 
{This argument is based on the observation that trapped surfaces
do not lie outside future event horizons. Fortunately for present 
purposes, this fact does not depend on asymptotic flatness, as is 
clear from the phrasing of Prop.~(12.2.4) of [17].}
we obtain
$$\langle {\tilde E}\rangle A(\partial {\cal B}(\tau))
\le 2\pi\chi(\partial {\cal B}(\tau))=4\pi(1-g)\quad ,\eqno{(19)}$$
for $\chi(\partial {\cal B}(\tau))$ the Euler characteristic and $g$ the
genus of $\partial {\cal B}(\tau)$.\footnote{$^3$}
{The factor of 2 in the middle term of (19) is missing in [16]. This 
error originates in their unnumbered equation preceding (9.7), the 
Gauss-Bonnet theorem, wherein the integrand should be the Gauss 
curvature, not the curvature scalar, the difference being the factor 
of 2.}
From this we recover the form of (15c), but with $\langle {\tilde E}
\rangle$ replacing $\langle E \rangle$. Notice these quantities differ 
by the additional $T_{ab}l^al^b$ factor in (17c) that does not appear
in ${\tilde E}$.

%\par\vfill\eject
\bigskip
\cen{\bf III. Discussion}
\smallskip
\noindent
We now consider a class of higher genus black holes, essentially 
those discussed in [3,4,18]. The metric
$$ds^2=V(r)dt^2-{1\over {V(r)}}dr^2 -r^2\left ( d\theta^2 
+\sinh^2 \theta d\phi^2\right )\eqno{(20)}$$
with 
$$V(r)=-1-{{\ell}\over 3}r^2-{{2m}\over r}+{{Z^2}\over {r^2}}
\eqno{(21)}$$
solves the Einstein-Maxwell system with electromagnetic potential
$A_0=Q/r$, $A_{\phi}=H\cosh\theta$, $A_r=A_{\theta}=0$, where $Q$ 
is the electric charge, $H$ the magnetic charge, and $Z^2=Q^2+H^2$. 
The non-zero components of the stress-energy tensor for these metrics 
can be expressed in the form
$$T_{\mu\nu}=\pm {{Z^2}\over {r^4}} g_{\mu\nu}\quad ,\eqno{(22)}$$
with the plus sign for $\mu,\nu=0,1$ and the minus sign for 
$\mu,\nu=2,3$, using the conventions of [15].

Horizons occur at roots $r=a>0$ of $V(r)$. Roots are easily located
by considering points of intersection of the curves
$$\eqalignno{
y_1=&-{{\ell}\over 3}r^4-r^2&{\rm (23a)}\cr
y_2=&2mr-Z^2&{\rm (23b)}\cr}$$
One fixes $y_1$ and varies the slope $2m$ and intercept $-Z^2$ of $y_2$
to move about in solution space. For $r>0$ generically there are either 
zero or two intercepts, and never three. There is a single intercept if 
either $Z=0$ and $m\ge 0$ or, more interestingly, $y_2$ is tangent to 
$y_1$. The latter case is an extreme solution and the point of intersection 
is a double root of $V(r)$. Extreme solutions minimize the mass for 
fixed $Z$ and, as we will see, realize the lower bound on horizon area.

Near a root $r=a$ of $V(r)$, we may expand
\def\Dr{\Delta r}
$$\eqalignno{
r=&a+\Dr&{\rm (24a)}\cr
V(r)=&\Dr V'(a)+{{\Dr^2}\over 2} V''(a)+{\cal O}(\Dr^3)
&{\rm (24b)}\cr}$$
If we replace $\Dr$ by a new variable $\sigma$ using
$$\sigma:=2\sqrt{{\Dr}\over {V'(a)}}\quad \Rightarrow \quad
d\sigma={{dr}\over {\sqrt{V'(a)\Dr}}}\quad ,\eqno{(25)}$$
then the metric near $r=a$ can be written as
$$ds^2\simeq\left [ {1\over 2}V'(r)\sigma\right ]^2 dt^2 -d\sigma^2
-a^2\left ( d\theta^2 +\sinh^2 \theta d\phi^2\right )\quad .\eqno{(26)}$$
If we treat $t$ as complex, then its imaginary part is a coordinate for 
a non-singular Euclidean submanifold iff it's periodic with period
$4\pi/V'(a)$. Then continuous Euclidean Green functions must have this
period, so by standard arguments the Hawking temperature is
$$T_H=-{{\hbar V'(a)}\over {4\pi k_B}}\quad .\eqno{(27)}$$
As with more familiar black holes, the Hawking temperature vanishes when
$V'(a)=0$; {\it i.e.}, when $r=a$ is a double root of $V(r)$. Hence the
extreme solutions above, and only the extreme solutions, are stable
against semi-classical decay by Hawking radiation; in particular, the
$m=Z=0$ black hole is unstable.

For solutions given by (20) and (21), by symmetry the area bound (15c)
on the horizon becomes a bound on the horizon ``radius'' $a$:
$$a^2\ge {{1}\over {-\langle E \rangle}}\quad .\eqno({(28)}$$
Since $T_{ab}$ is constant over the horizon, $\langle E \rangle$ is 
easily computed from (22) to give:
$$\langle E \rangle=\ell +{{Z^2}\over {a^4}}\quad ,\eqno{(29)}$$
Plugging this into (28) and rearranging, we obtain
$$0\le -\ell a^4 - a^2 -Z^2\quad ,\eqno{(30)}$$
which is saturated iff (recall $\ell<0$ here)
$$a^2={{1+\sqrt{1-4\ell Z^2}}\over {-2\ell}}\quad .\eqno{(31)}$$
But this is the simultaneous root of $V(r)=V'(r)=0$, so (31) is 
satisfied iff the metric of form (20, 21) is an extreme solution
iff it has zero temperature.

The area bound may point to interesting features of the mechanical laws 
governing asymptotically locally anti-de Sitter black holes. Consider the
semi-classical decay of the horizon by Hawking radiation. If the flux at
infinity is positive and if the surface gravity $k$ is positive, then the 
first law of black hole thermodynamics, $\delta M=k \delta A$, implies
that this process will cause the horizon area to shrink. The horizon area
$A(t)$ will be a decreasing function of time, bounded below by a positive 
number whenever $g>1$. Thus, $A(t)$ will converge to a positive value, and
the spacetime will approach a steady state, just as asymptotically flat 
charged black hole spacetimes do (but without any restored supersymmetries). 
This is reminiscent of the conjecture of Horowitz and Myers [9], argued 
from an entirely different point of view, that the mass-energy of certain 
5-dimensional spacetimes with negative scalar curvature is bounded below, 
even though the positive energy theorem fails for these spacetimes. Gibbons 
[8] has recently used the mean curvature flow method of Geroch [19] to argue 
that such a lower bound applies for mass-energy of 4-dimensional asymptotically 
locally adS black holes on any hypersurface that is a moment of time symmetry.

In certain circumstances, the horizon could still ``decay'' by first 
{\it growing} until it develops self-intersections. This can result 
in a decrease in the genus, resulting in turn in a decrease in the 
area bound (this process can sometimes drive the genus and, hence, 
the area bound, to zero). So-called temporarily toroidal horizons 
arising in numerical simulations of asymptotically flat spacetimes 
change genus in this manner, although the effect depends on a choice 
of spatial slicing---there are slicings in which these horizons are 
never toroidal (see [2] for details and further references; also [20]). 
This process, however, is not available to all non-zero genus horizons. 
Indeed, for the black holes discussed above, the horizon and the boundary 
at infinity are linked in such a manner that  horizon topology change is 
precluded.

\bigskip
\cen{\bf Acknowledgements}
\smallskip
\noindent
I thank Greg Galloway for several discussions and for calling to 
my attention several references related to Hawking's horizon 
topology theorem. I thank Robb Mann for discussions of locally adS 
black holes. This work was partially supported by the Natural Sciences 
and Engineering Research Council of Canada.

\bigskip
\cen{\bf References}
\smallskip
\noindent
\item{[1]}{J.L.~Friedman, K.~Schleich, and D.M.~Witt, 
{\it Phys. Rev. Lett.}~{\bf 71} (1993), 1486,
gr-qc/9305017; Erratum-ibid.~{\bf 75} (1995),1872.}
\item{[2]}{G.J.~Galloway, K.~Schleich, D.M.~Witt, and E.~Woolgar, 
preprint (1999), gr-qc/9902061.}
\item{[3]}{R.B.~Mann, {\it Class.~Quantum Gravit.}~{\bf 14} (1997), 
2927, gr-qc/9705007.}
\item{[4]}{D.R.~Brill, J.~Louko, and P.~Peldan, 
{\it Phys.~Rev.}~D{\bf 56} (1997), 3600,gr-qc/9705012.}
\item{[5]}{L.~Vanzo, {\it Phys.~Rev.}~D{\bf 56} (1997), 6475.}
\item{[6]}{C.M.~Will, {\it Theory and experiment in gravitational
physics} (Cambridge University Press, Cambridge, 1981), p.~82.}
\item{[7]}{C.M.~Hull, {\it Commun.~Math.~Phys.}~{\bf 90} (1983), 545.}
\item{[8]}{G.W.~Gibbons, {\it Class.~Quantum Gravit.}~{\bf 16} (1999), 
1677.}
\item{[9]}{G.T.~Horowitz and R.C.~Myers,
{\it Phys.~Lett.}~B{\bf 428} (1998), 297, hep-th/9803066.}
\item{[10]}{Y.~Shen and S.~Zhu, {\it Math.~Ann.}~{\bf 309} (1997), 107}
\item{[11]}{G.J.~Galloway, {\it Contemp.~Math.}~{\bf 170} (1994), 113.}
\item{[12]}{S.W.~Hawking, in {\it Black Holes}, eds.~C.~DeWitt and 
B.~DeWitt (Gordon and Breach, New York, 1973), 1.}
\item{[13]}{S.A.~Hayward, T.~Shiromizu, and K.~Nakao,
{\it Phys.~Rev.}~D{\bf 49} (1994), 5080, gr-qc/9309004.}
\item{[14]}{P.T.~Chru\'sciel and G.J.~Galloway, preprint (1996), 
gr-qc/9611032.}
\item{[15]}{R.~Penrose and W.~Rindler, {\it Spinors and space-time}
Vol.~I (Cambridge University Press, Cambridge, 1984).}
\item{[16]}{S.W.~Hawking and G.F.R.~Ellis, {\it The large scale structure 
of space-time} (Cambridge University Press, Cambridge, 1973).}
\item{[17]}{R.M.~Wald, {\it General Relativity} (University of 
Chicago Pressm Chicago, 1984).}
\item{[18]}{R.B.~Mann, in {\it Internal Structure of Black Holes
and Spacetime Singularities}, eds.~L. Burko and A.~Ori, 
{\it Ann.~Israeli Phys.~Soc.}~{\bf 13} (1998), 311.}
\item{[19]}{R.~Geroch, {\it Ann.~N.Y.~Acad.~Sci.}~{\bf 224} (1973), 108.}
\item{[20]}{T.~Jacobson and S.~Venkataramani, {\it Class.~Quantum
Gravit.}~{\bf 12} (1995), 1055.}
\bye